\documentclass[
aps,%
12pt,%
final,%
notitlepage,%
oneside,%
onecolumn,%
nobibnotes,%
nofootinbib,
superscriptaddress,%
noshowpacs,%
centertags]%
{revtex4}

\usepackage{graphicx}

\begin{document}
\title{Prospects for strangeness measurement in ALICE}
\author{\firstname{R.} \surname{Vernet}}
\email[]{renaud.vernet@cern.ch}
\affiliation{Consorzio COMETA, Catania, Italy}
\collaboration{ALICE}
\begin{abstract}
  The study of strangeness production at LHC will bring significant information on the bulk chemical properties, its dynamics and the hadronisation mechanisms involved at these energies. 
  The ALICE experiment will measure strange particles from topology (secondary vertices)
  and from resonance decays over a wide range in transverse momentum and shed light on this new QCD regime.
  These motivations will be presented as well as the identification performance of ALICE for strange hadrons.
\end{abstract}

\maketitle
\section{Introduction}

From the beginning of the search for the Quark-Gluon Plasma (QGP), strangeness has been a powerful probe of the chemical and dynamical properties of the medium created in heavy-ion collisions.
An enhanced production of strangeness was originally considered as a main signature of a hot and thermalised system having partonic degrees of freedom. Indeed, the creation of strange quark-antiquark pairs in a medium in which quarks and gluons are deconfined should be larger than in a pure hadronic system, which should result in an enhancement of strange particles: the more strangeness they contain, the greater the enhancement~\cite{Rafelski:1982pu}. In addition, because the volume involved in heavy-ion collisions is much bigger than in small systems, a statistical relative enhancement of strange particles is also expected~\cite{Tounsi:2001ck}.

The statistical hadron-resonance gas models at equilibrium, as well as Wroblewski factor predictions, describe successfully the strange particle ratios at SPS and RHIC~\cite{Andronic:2005yp,Braun-Munzinger:2001as}, which indicates a chemical equilibrium has been possibly reached, which a QGP may be the vector of.
Whereas statistically understood, hadron production itself and its mechanisms still remain an open question.
Several so-called `recombination' models argue that two competitive mechanisms govern hadron production: low- and intermediate-momentum hadrons are mainly created via coalescence of `soft' quarks, while high momenta are ruled by fragmentation~\cite{Hwa:2006vb,Fries:2003kq,Fries:2003fr,Greco:2003xt}.
This idea is confirmed experimentally by several results that reveal a strong discrepancy between meson and baryon production when considering observables such as momentum distributions or nuclear modification factors ($R_{CP}$)~\cite{Adams:2006wk,Hippolyte:2006ra,Adams:2005dq}. The understanding of hadron production at the Large Hadron Collider (LHC) will imply, therefore, the understanding of the relative contribution of these competing mechanisms and how they interplay in the intermediate momentum region, where the baryon-meson discrepancy is more evident.

It is hence mandatory to measure hadrons at LHC in a region of momentum as wide as possible to understand the underlying QCD processes.
Strange secondary vertices and resonances offer the opportunity to do so, since their identification methods (topology and invariant mass, respectively) may be limited by statistics only. Besides, resonances provide an additional possibility to investigate the collision dynamics at its early times, in particular for probing the duration between chemical and thermal freeze-outs.

While \mbox{Pb-Pb} collisions will provide exceptional conditions to study the QGP due to large volume and lifetime of the system, LHC's first data will be \mbox{p-p} collisions, which will allow to probe values of Bjorken-$x$ never reached so far. 
In addition, \mbox{p-p} collisions will not only serve as benchmark for \mbox{Pb-Pb} physics, but will also be a way to test pQCD, which is of fundamental interest in this new energy regime.
In that respect, the preparation of the tools to measure strange particles at LHC is of major importance, either in \mbox{Pb-Pb} or \mbox{p-p} colliding systems.
ALICE (A Large Ion Collider Experiment) is specifically designed for the study of the QGP at LHC.
Thanks to its large acceptance and its highly-precise tracking apparatus, ALICE will satisfy the need of identifying strange particles in a range of $p_T$ covering the soft, intermediate and hard regimes.

In the two following sections, we present ALICE capabilities to measure strange i) secondary vertices (with a particular emphasis on hyperons) and ii) resonances, considering the particles $K^0(892)$ and $\phi(1020)$. We describe the methods developed for the detection of these particles and show their expected reconstruction efficiencies and yields, and give a prospect of what can be achievable within the first \mbox{Pb-Pb} and \mbox{p-p} runs.

\section{Identification of secondary vertices}

\begin{figure*}[t!]
  \includegraphics[scale=0.25]{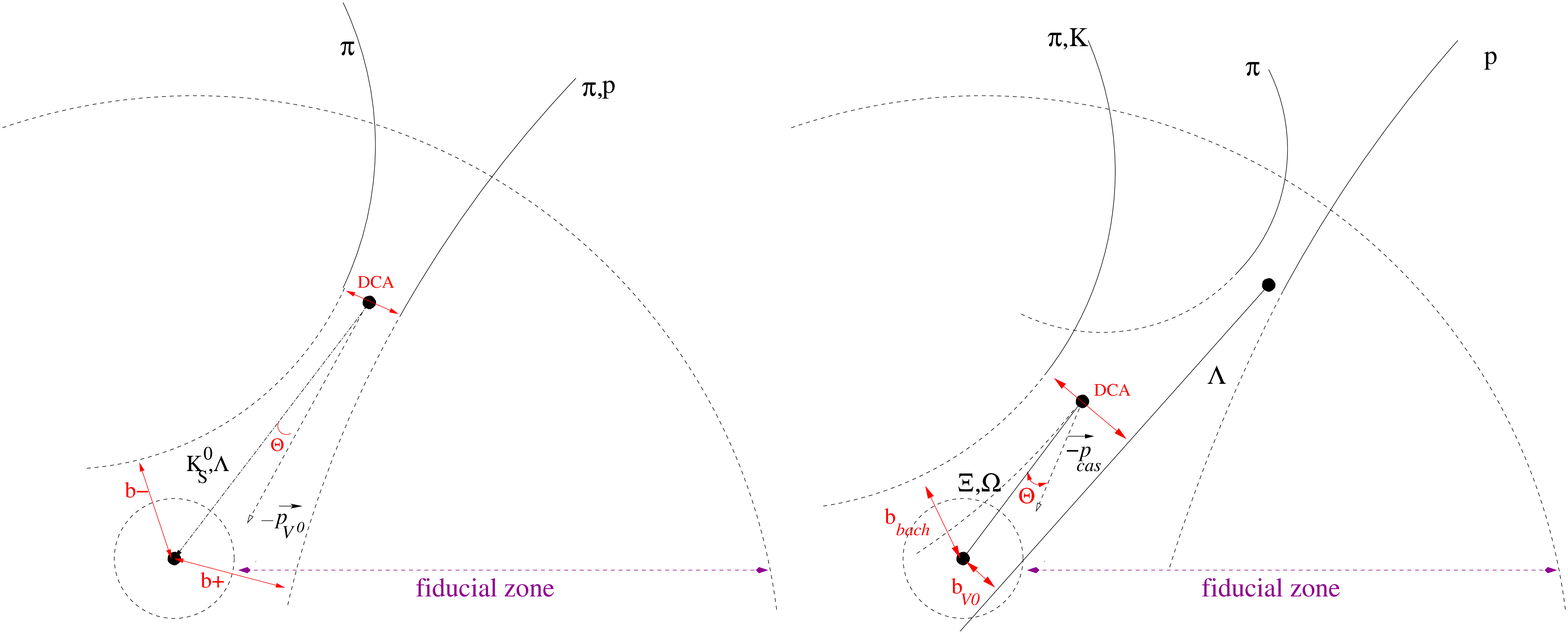}
  \caption{\label{fig:topologicalSelections}${\rm V}^0$ (left) and cascade (right) topological selections and fiducial zone.}
\end{figure*}

Strange particles $K^0_S$, $\Lambda$, $\Xi$ and $\Omega$ decay, via weak interaction, few centimeters away from the primary vertex. Therefore, their charged decay modes may be identified using topological methods that consist in selecting daughter track candidates according to geometrical criteria, as shown in Figure~\ref{fig:topologicalSelections}.
The identification of ${\rm V}^0$s, namely $K^0_S \rightarrow \pi^+\pi^-$ and $\Lambda \rightarrow p\pi^-$, is done applying geometrical selections on the daughter impact parameters $(b^+,b^-)$, the distance of closest approach between the daughters ($DCA$) and the pointing angle.
A condition on the $\rm{V}^0$ decay position is also imposed by means of a `fiducial zone', defined by two extreme radii in the transverse plane.

The identification of cascades ($\Xi^- \rightarrow \Lambda \pi^-$ and $\Omega^- \rightarrow \Lambda K^-$) consists in associating a ${\rm V}^0$ candidate with a single-track candidate (the `bachelor'), using selections on the ${\rm V}^0$ mass and impact parameter, the $DCA$ between the ${\rm V}^0$ and the bachelor, the bachelor impact parameter, and the cascade pointing angle and transverse decay radius.
The detailed procedure can be found in reference~\cite{Vernet:2006alicenote}.

Single particle identification (PID) is not required in these methods, which makes them especially efficient at intermediate and high transverse momenta, where energy loss and time of flight identification fail.
As a matter of fact, the studies presented in this section were obtained using the ALICE tracking devices TPC and ITS, without any requirement on single-track energy loss inside these detectors.

The two following subsections present respectively the secondary-vertex identification performance of ALICE for \mbox{Pb-Pb} and \mbox{p-p} colliding systems.

\subsection{Hyperon reconstruction in \mbox{Pb-Pb} at $\sqrt{s_{NN}}=5.5~TeV$}

A precise measurement of particle production in the \mbox{Pb-Pb} system is very challenging given the very high track density that implies a large combinatorial contamination.
For this reason, the track topological selection must be fine-tuned in order to get as many secondary vertex candidates as possible, keeping the background at a reasonable level, and allowing thus to distinguish within the very first minutes of data taking the signal from the background in their invariant mass spectrum.

The results shown in this section were obtained from events simulated with the HIJING generator~\cite{Wang:1991ht} parameterised for a charged particle density (${\rm d}N_{ch}/{\rm d}y$) at mid-rapidity of 4000. The $\Lambda$, $\Xi$ and $\Omega$ (and their antiparticles) multiplicities at mid-rapidity and the inverse slopes of their exponential $p_T$ spectra are detailed in reference~\cite{Vernet:2006alicenote}.

\begin{figure*}[t!]
  \includegraphics[scale=1.7]{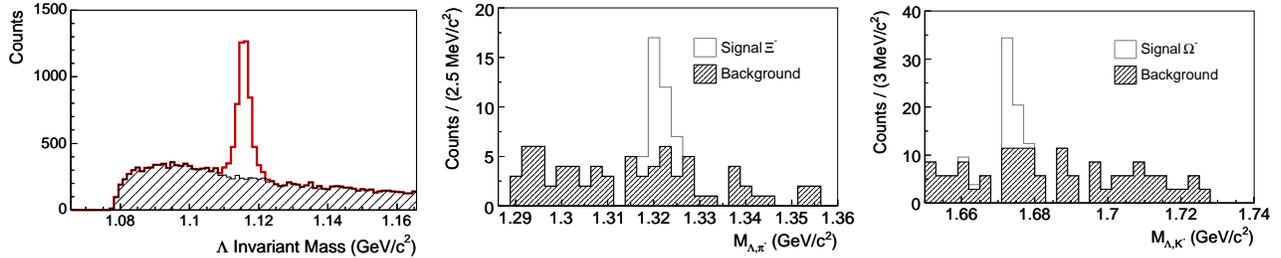}
  \caption{\label{fig:specMassHyp}Reconstructed invariant mass spectra of $\Lambda$ (left), $\Xi$ (middle) and $\Omega$ (right) obtained from central HIJING \mbox{Pb-Pb} events at $\sqrt {s_{NN}}=5.5~TeV$.}
\end{figure*}

Figure~\ref{fig:specMassHyp} shows the invariant mass spectra of $\Lambda$ and $\Xi$ obtained after processing the full reconstruction chain on 300 of the aforementioned events. 
For the $\Omega$, a dedicated study was performed due to the low efficiency and requested 6000 events.
One can observe that the signal is clearly visible for all three hyperons, with signal over background ratios ($S/B$) close to the unity and resolutions on invariant mass around $3~MeV/c^2$. 
The average overall efficiency for $\Lambda$ ($\Xi$, $\Omega$) reconstruction corresponds to 11\% (0.45\%, 0.36\%). In other words, one should expect an mean number of 11 (0.07, 0.01) identified $\Lambda$ ($\Xi$, $\Omega$) per event.

To draw an estimate of the capability of ALICE to identify high-$p_T$ hyperons, we use the overall reconstruction efficiency as function of the hyperon $p_T$, which we multiply by the $p_T$ spectrum expected in \mbox{Pb-Pb} collisions.
Each $p_T$ spectrum is obtained assuming the statistics of $10^7$ central events estimated for the first \mbox{Pb-Pb} run and considering two extreme values of inverse slopes.

\begin{figure*}[!t]
  \includegraphics[scale=1.7]{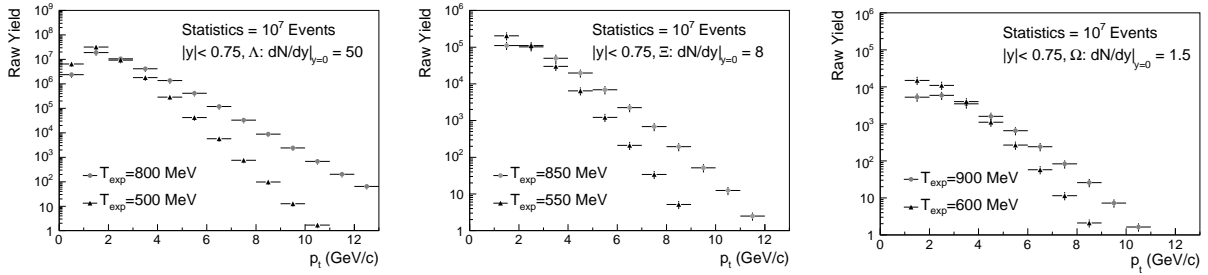}
  \caption{\label{fig:hyperonRawYields}$\Lambda$ (left), $\Xi$ (middle) and $\Omega$ (right) reconstructed yields expected for $10^7$ Pb-Pb central events at $\sqrt {s_{NN}}=5.5~TeV$. For each particle two inverse slopes of the exponential $p_T$ spectrum are considered. They are reported on each figure together with the assumed multiplicity per unit of rapidity.}
\end{figure*}

The reconstructed yield spectra as function of $p_T$ are presented in Figure~\ref{fig:hyperonRawYields}.
They show that a statistics of $\sim 100$ $\Lambda$ ($\Xi$, $\Omega$) can be expected at $p_T$ as high as 10 (8, 6) $GeV/c$ within the first run. 
In order to take into account hard processes, one could consider an additional power-law contribution in the high-$p_T$ part of the generated distributions; that would push these statistical limits to higher values.

\subsection{Hyperon reconstruction in \mbox{p-p} at $\sqrt{s_{pp}}=14~TeV$}

The identification of secondary vertices in p-p collisions is rather different from that in Pb-Pb.
Since in the small system particle multiplicities are low, combinatorial background is even lower, which allows therefore topological selections to be loosened in order to gather more signal.
However, the \mbox{p-p} system is affected by a larger error on primary vertex position measurement in low-multiplicity events, which can substantially alter the reconstruction efficiency.
A detailed study on these effects can be found in reference~\cite{Gaudichet:2005alicenote}.

\begin{figure*}[t!]
  \includegraphics[scale=1.7]{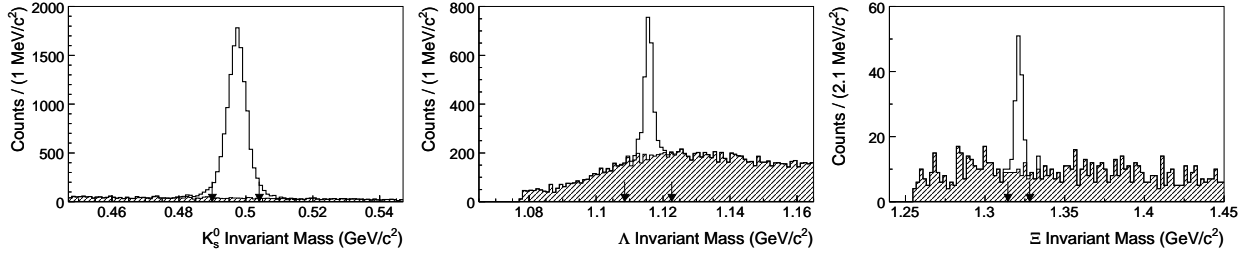}
  \caption{\label{fig:ppmass}Reconstructed invariant mass spectra of $K^0_S$ (left), $\Lambda$ (middle) and $\Xi$ (right) obtained from \mbox{p-p} collisions at $\sqrt{s_{pp}}=14~TeV$.}
\end{figure*}

\begin{figure*}[t!]
  \includegraphics[scale=1.7]{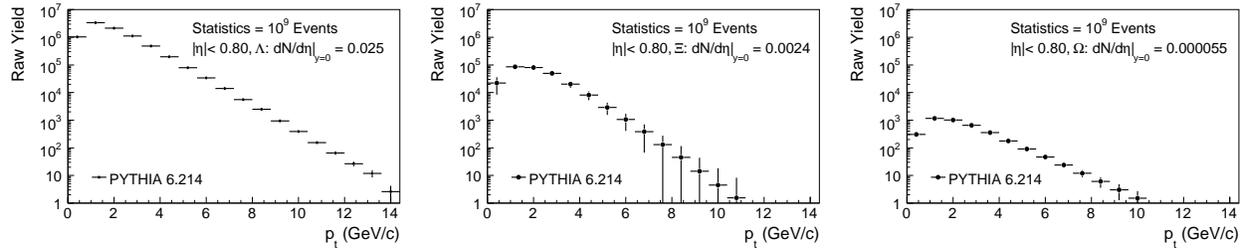}
  \caption{\label{fig:ppyields}$\Lambda$ (left), $\Xi$ (middle) and $\Omega$ (right) reconstructed yields expected for $10^9$ \mbox{p-p} events at $\sqrt {s_{pp}}=14~TeV$. The assumed multiplicities per unit of rapidity are reported on each figure.}
\end{figure*}

Figure~\ref{fig:ppmass} shows the $K^0_S$, $\Lambda$ and $\Xi$ reconstructed invariant mass spectra obtained from \mbox{p-p} events generated with PYTHIA6.214~\cite{Sjostrand:2001yu}.
The $S/B$ ratio is very high for the $K^0_S$ and close to unity for $\Lambda$ and $\Xi$.
The obtained precision on invariant mass measurement is about $5~MeV/c^2$ for $K^0_S$ and is comparable to that of \mbox{Pb-Pb} for what concerns $\Lambda$ and $\Xi$.
An estimate of the $\Lambda$ and $\Xi$ reconstructed yields has been performed as well.
They are illustrated in Figure~\ref{fig:ppyields} as a function of $p_T$ in the rapidity range $|y|<0.8$, assuming a number of $10^9$ events. 
The highest hyperon $p_T$ reached (considering a statistics of $\sim 100$ entries in the corresponding bin) within the first \mbox{p-p} run is also comparable to that obtained in \mbox{Pb-Pb}.

\section{Strange resonance identification.}

The results shown in this section concern the $K^0(892)$ and $\phi(1020)$ resonances. 
For the sake of simplicity, they are indicated respectively by $K^*$ and $\phi$. 
The decay modes investigated in this section are $K^*\rightarrow K^+\pi^-$ (or $\overline{K^*}\rightarrow K^-\pi^+$) and $\phi\rightarrow K^+K^-$.
Since resonances decay very early, their decay daughters are not discernible from other primary particles. 
Resonances are identified via invariant mass reconstruction methods that combine all possible pairs of primary daughter candidates.
The resulting background being very high since no selection other than PID or track quality is applied, we estimate it by means of `like-sign' or `event mixing' procedures.
Both aim to reconstruct only non-signal candidates and to reproduce the background shape.
The kaon and pion identification used in these studies rely on a combined PID obtained from the energy loss (${\rm d}E/{\rm d}X$) in the TPC, and from the time measurement of the TOF. The analyses have been performed on a sample of 1.5M \mbox{p-p} events obtained with the particle generator PYTHIA6.214 at the energy $\sqrt{s_{pp}}=14~\rm{TeV}$.

\subsection{$K^0(892)$ identification in \mbox{p-p} at $\sqrt{s_{pp}}=14~TeV$}

\begin{figure*}[t!]
  \includegraphics[scale=0.8]{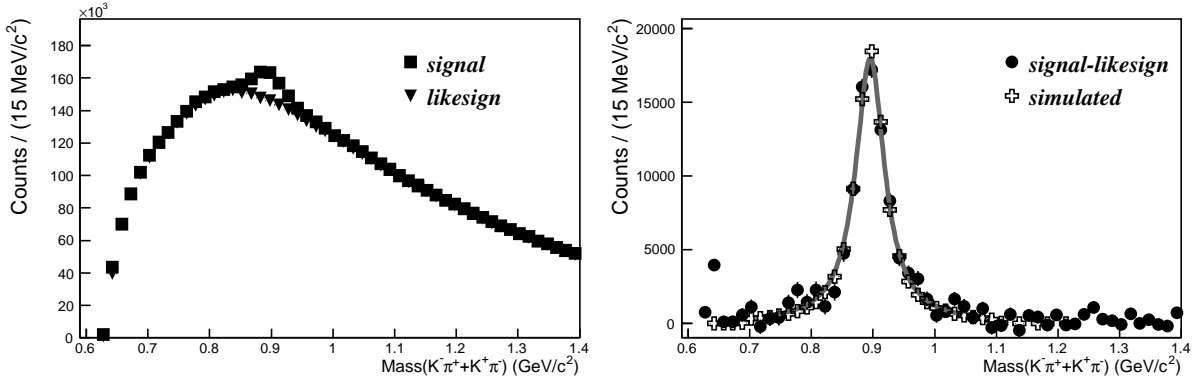}
  \caption{\label{fig:KstarOverall}Left: $K^*$ invariant mass spectrum (unlike-sign pairs, squares) and estimated background (like-sign pairs, triangles). Right: $K^*$ invariant mass spectrum after background subtraction (circles) and true signal (crosses). The curve is a Breit-Wigner fit. $p_T$ and $y$ are integrated in both figures.}
\end{figure*}

We present here the results obtained for the identification of $K^*$.
Figure~\ref{fig:KstarOverall} shows, on the left part, the reconstructed invariant mass spectrum of the `unlike-sign' pairs $(K^+\pi^-)$ and $(K^-\pi^+)$ together with the estimated `like-sign' background.
The like-sign background is in a good agreement with the unlike-sign spectrum, apart of course in the region of the $K^*$ mass where the signal peak is clearly visible.
The right part of the figure shows the resulting subtraction between the unlike- and like-sign spectra, where the associated simulated $K^*$ spectrum is superimposed. 
A Breit-Wigner fit of the subtracted spectrum returns a mean mass of $895~\rm {MeV/c^2}$ and a width of $56~MeV/c^2$, to be compared to the expected values 896 and $52~MeV/c^2$~\cite{Yao:2006px}. The agreement is good for the mass calculation, whereas less accurate for the width, due to detector-resolution effects.
The $K^*$ yield is calculated here from the fit integral; the relative discrepancy with the true signal is 0.9\%, which indicates the background is estimated accurately.

\begin{figure*}[t!]
  \includegraphics[scale=0.8]{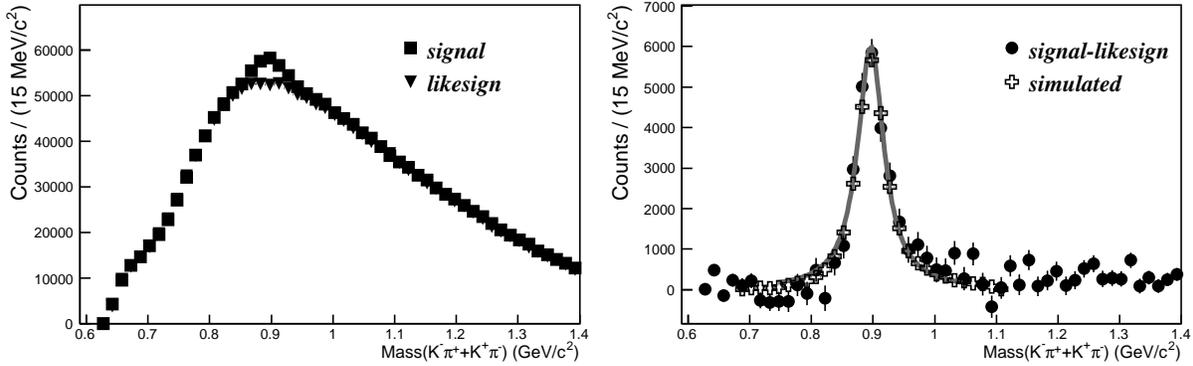}
  \caption{\label{fig:KstarLowPt}Left: $K^*$ invariant mass spectrum (unlike-sign pairs, squares) and estimated background (like-sign pairs, triangles) for $p_T<0.5~GeV/c$. Right: $K^*$ invariant mass spectrum after background subtraction (circles) and true signal (crosses). The curve is a Breit-Wigner fit.}
\end{figure*}

\begin{figure*}[t!]
  \includegraphics[scale=0.8]{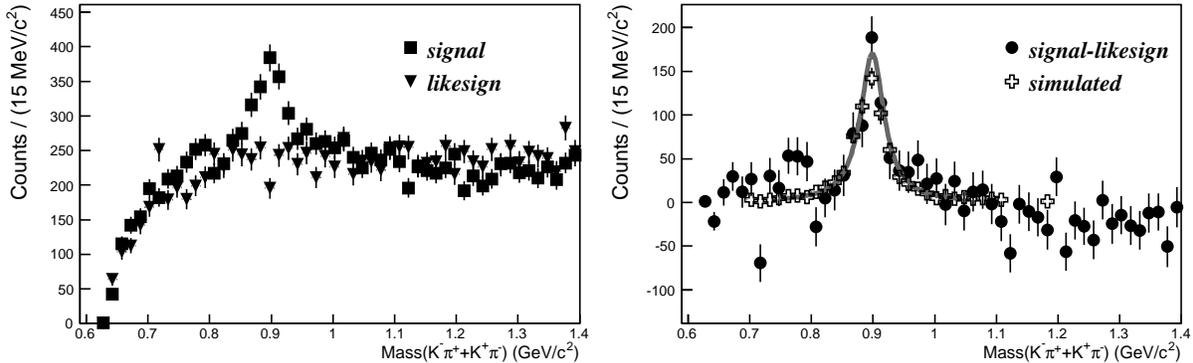}
  \caption{\label{fig:KstarHighPt}Left: $K^*$ invariant mass spectrum (unlike-sign pairs, squares) and estimated background (like-sign pairs, triangles) for $3.5<p_T<4~GeV/c$. Right: $K^*$ invariant mass spectrum after background subtraction (circles) and true signal (crosses). The curve is a Breit-Wigner fit.}
\end{figure*}

\begin{figure*}[t!]
  \includegraphics[scale=0.6]{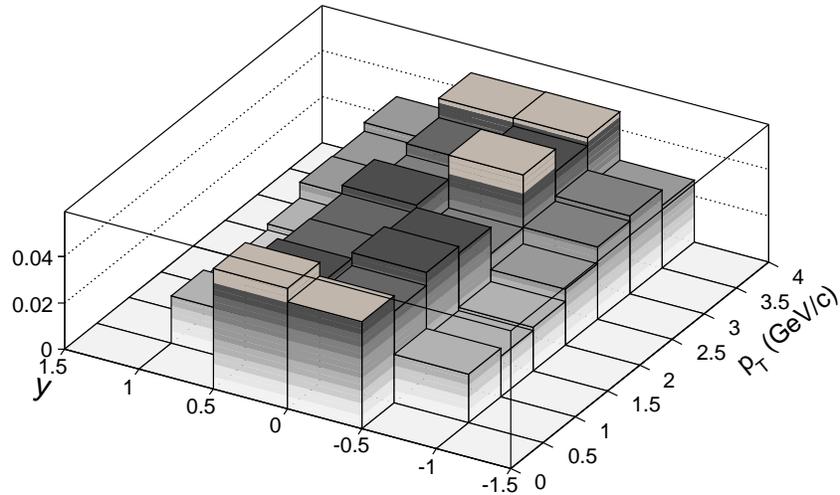}
  \caption{\label{fig:matrixKstar}$K^*$ reconstruction efficiency as a function of transverse momentum ($p_T$) and rapidity ($y$) in \mbox{p-p} collisions at $\sqrt{s_{pp}}=14~TeV$.}
\end{figure*}

A similar signal-calculation procedure was applied for various bins in $p_T$. Figures~\ref{fig:KstarLowPt} and \ref{fig:KstarHighPt} show the invariant mass plots for two selected $p_T$ intervals: $[0-0.5]$ and $[3.5-4]~GeV/c$.
In each bin of $p_T$, the background estimation is in reasonable agreement with the unlike-sign spectrum. This agreement does not seem to be affected, within the fluctuation amplitude, by the fact that the spectrum shapes in the two bins strongly differ.
Furthermore, despite the relative low statistics resulting in each bin, the signal extraction in both cases is still possible.
We hence chose to draw a first estimate of the overall reconstruction efficiency as a function of $p_T$ and $y$. 
The corresponding results are presented in Figure~\ref{fig:matrixKstar}, focusing on the intervals $[p_T \times y] = [(0\rightarrow 4~GeV/c) \times (-1.5\rightarrow 1.5)]$. 
The average efficiency is about 4\% in the most central bins in rapidity, but drops substantially ($\sim 2\%$) at larger $|y|$, due to the lower single-track reconstruction efficiency in that region.

\subsection{$\phi(1020)$ identification}

\begin{figure*}[t!]
  \includegraphics[scale=0.38]{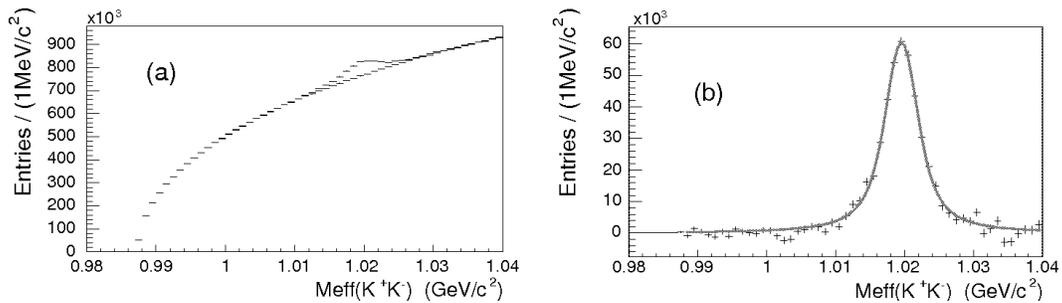}
  \caption{\label{fig:phiInvMassPbPb}Left: $\phi$ invariant mass spectrum and estimated background (like-sign pairs) for $p_T>2.2~GeV/c$. Right: $\phi$ invariant mass spectrum after background subtraction. The curve is a Breit-Wigner fit.}
\end{figure*}

$\phi$ identification in ALICE was already investigated in Pb-Pb collisions~\cite{DeCaro:2003alicenote}. 
This study, performed considering a multiplicity of ${\rm d}N_{ch}/{\rm d}y=6000$ using the HIJING generator, illustrates the fact that the identification of the $\phi$ resonance is possible in spite of the very large combinatorial background.
Indeed, Figure~\ref{fig:phiInvMassPbPb} shows accurate calculation of the like-sign background and signal extraction for $p_T(\phi)>2.2~\rm{GeV/c}$ using a single-track combined PID from TPC and TOF. 
The obtained values for the reconstructed mass ($1019.60~\rm{GeV/c^2}$) and width ($4.32~\rm{MeV/c^2}$) are compatible with PDG values~\cite{Yao:2006px}.

\begin{figure*}[t!]
  \includegraphics[scale=0.8]{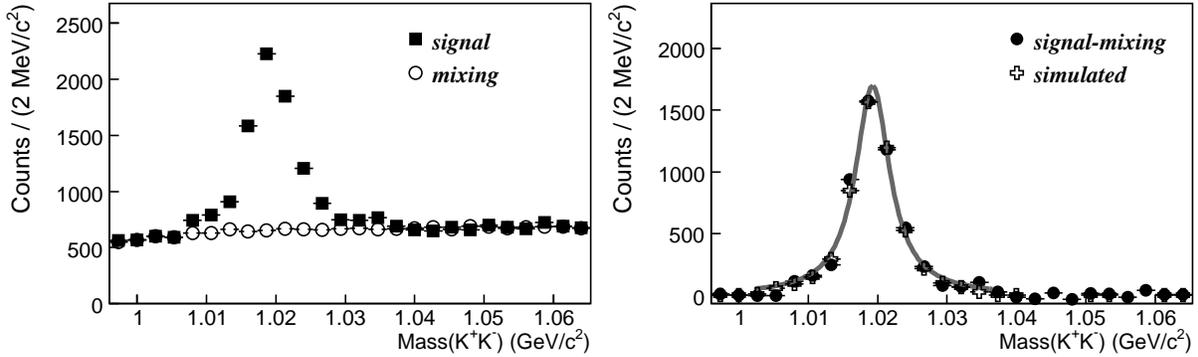}
  \caption{\label{fig:phiInvMassPP}Left: $\phi$ invariant mass spectrum (unlike-sign pairs, squares) and estimated background (mixed pairs, circles). Right: $\phi$ invariant mass spectrum after background subtraction (circles) and true signal (crosses). The curve is a Breit-Wigner fit.}
\end{figure*}

We now wish to focus on $\phi$ identification in \mbox{p-p} at $\sqrt{s_{pp}}=14~TeV$. 
Single particle PID comes from TPC and TOF. We keep the kaon-candidate tracks that satisfy the two following conditions: i) $p_K > 0.4$ and ii) $p_K > p_e$, $p_\mu$, $p_\pi$, $p_p$ where $p_X$ represents the probability for a track to be an electron, a muon, a pion, a kaon or a proton.
The invariant mass spectra obtained from all primary $K^+K^-$ associations satisfying the condition $0<p_T(\phi)<4~\rm{GeV/c}$ and $-1.5<y(\phi)<1.5$ , are displayed in Figure~\ref{fig:phiInvMassPP}.
The background was estimated via event-mixing procedure and then normalised. One can observe that the mixed background is in good agreement with the real background except in zones where correlations are expected.
The measured mass and width obtained from a Breit-Wigner fit are respectively 1019 and 5.53~$MeV/c^2$. 
The mass is compatible with the expected value 1019~$MeV/c^2$, but the width measurement is not as precise (to be compared with $4.26~MeV/c^2$) because of detector resolution.
The yield calculation using the fit integral reveals a relative discrepancy of $\sim 2.5\%$ with respect to the number of true $\phi$ actually found, which means the calculation is rather accurate. The overall reconstruction rate, regardless of the $p_T$ and $y$ of the $\phi$, is about $2.8\%$. 
Furthermore, we expect the $\phi$ to be identified at least up to $p_T\sim 4~GeV/c$ given that the studies on $K^*$ (presented in the previous section) do not reveal any major difficulty to detect kaons in the corresponding momentum range (in which TOF is still efficient). This assumption seems to be confirmed by preliminary studies.

\section{Conclusions}
In these proceedings we have discussed the importance of measuring strange secondary vertices and resonances in a broad range of transverse momentum, and have shown how ALICE will face this challenge thanks to its tracking apparatus and identification methods developed for that purpose.
From the studies shown here, we conclude that ALICE is perfectly suited for the measurement of the strange secondary vertices $K^0_S$, $\Lambda$, $\Xi$ and $\Omega$ in both \mbox{Pb-Pb} and \mbox{p-p} colliding systems. Topological selections have been tuned to get a good compromise between signal and background with a precision on reconstructed invariant masses of a few $MeV/c^2$ and overall efficiencies of $\sim 11\%$ for $\rm{V^0s}$ and $\sim 0.5\%$ for cascades. We have shown that PID is not mandatory for this measurement, and that these particles can be detected within the very first minutes of LHC run. With statistics as high as the ones expected for the first runs, they should be identified in a range of $p_T$ varying from almost 0 to values of at least $10~GeV/c$.
We have also shown that strange resonances can be measured via invariant mass methods, and that the background can be calculated accurately with both like-sign and event-mixing estimators. Concerning the $K^*$, reaching $p_T$'s as high as $4~GeV/c$ is not problematic, and so it should be for the $\phi$. However, the identification of higher-$p_T$ resonances should be done without using PID.

The tools to identify strange secondary vertices and resonances are ready for the upcoming data in 2008, and we infer they can provide `first-physics' observables.
Within a larger time scale, the statistics of strange particles reconstructed with ALICE will by far overstep that of previous experiments, and will allow several new studies that were barely achievable up to now because of statistics, such as strange high-$p_T$ hadron quenching or strange hard-soft correlations. We believe these fields of research will be among the most exciting ones in strangeness physics.

\newpage

\bibliography{refs}

\end{document}